# Ultra-low Energy charge trap flash based synapse enabled by parasitic leakage mitigation

Shalini Shrivastava*, Tanmay Chavan and Udayan Ganguly†
Department of Electrical Engineering, Indian Institute of Technology Bombay, Mumbai, 400076, India

*shalinishrivastava@iitb.ac.in; †udayan@ee.iitb.ac.in

**Abstract**

Brain-inspired computation promises complex cognitive tasks at biological energy efficiencies. The brain contains $10^4$ synapses per neuron. Hence, ultra-low energy, high-density synapses are needed for spiking neural networks (SNN). In this paper, we use tunneling enabled CTF (Charge Trap Flash) stack for ultra-low-energy operation (1F); Further, CTF on an SOI platform and back-to-back connected pn diode and Zener diode (2D) prevent parasitic leakage to preserve energy advantage in array operation. A bulk $100\ \mu m \times 100\ \mu m$ CTF operation offers tunable, gradual conductance change ($\Delta G$) i.e. $10^4$ levels, which gives $100\times$ improvement over literature. SPICE simulations of 1F2D synapse shows ultra-low energy ($\leq 3$ fJ/pulse) at 180nm node for long-term potentiation (LTP) and depression (LTD), which is comparable to energy estimate in biological synapses (10 fJ). A record low learning rate (i.e., maximum $\Delta G$<1% of $G$-range) is observed – which is tunable. Excellent reliability ($> 10^6$ endurance cycles at full conductance swing) is observed. Such a highly energy efficient synapse with tunable learning rate on the CMOS platform is a key enabler for the human-brain-scale systems.

The high-performance computing such as IBM Watson supercomputer is $10^6 \times$ less energy- and $10^3 \times$ less area-efficient than biological neural network [1]. Each biological pre-neuron transmits information to post-neurons through a synapse (Fig. 1a). There are approximately $10^4$ synapses per neuron, which make the synapses the largest component of the neural network. Hence, area density and energy of operation of the synapse are critical determinants of system performance. Recently, instead of large digital circuits to represent analog weights [2], various nanoscale memristive synapses have been proposed [3-16]. These memristive devices store the weight as an analog conductance ($G$) value to provide excellent areal density improvement [17]. The synapse "learns" by conductance change ($\Delta G$) which depends upon the spike time difference ($\Delta t$) of pre- and post-neurons (Fig. 1b), known as spike time dependent plasticity (STDP). Long-term potentiation (LTP: $\Delta G > 0$) and depression (LTD: $\Delta G < 0$) is needed. For memristive synapses, the application of a $\Delta t$-dependent pulse voltage ($V_{pulse}$) at fixed pulse-width ($t_p$) causes $\Delta G$. The $\Delta G$ per pulse depends upon both the instantaneous conductance $G$ and $V_{pulse}$. For varying $V_{pulse}$, $\Delta G$ increases/decreases with $V_{pulse}$ (Fig. 1c), which is widely shown [13][18]. In addition, for fixed $V_{pulse}$ applied repeatedly, $G$ increases/decreases gradually and then saturates (Fig. 1d). For fixed $V_{pulse}$, the maximum $\Delta G$ (i.e., $\Delta G^{max}$) occurs initially. Synapses have two key challenges. First, both LTP and LTD need to be gradual for SNN algorithms in software [19]. In fact, the learning rate, i.e., $\Delta G^{max}$ for repeated *maximum* $V_{pulse}$ needs to be low (<2% of the total range of $G$) for stable weight evolution in a network during training [20-21][27]. In other words, the change in G should saturate after a larger number of identical $V_{pulses}$ i.e., in excess of 256 identical pulses for analog-valued datasets like Fischer Iris [22]. Second, low energy operation (especially write-energy) in the analog synapse is a critical challenge for memristors [17]. Presently, nanoscale memristive synapse may not have gradual LTP and LTD. For example, some synapses are binary (e.g., $HfO_2$ RRAM [23]). Other synapse has gradual LTP but abrupt LTD (e.g., Phase Change Memory), which requires novel synapse circuit design with enhanced controller complexity [24-25]. Other memristors (e.g., Mn doped $HfO_2$ [12] or PCMO RRAM [13]) have analog LTP and LTD with <100 states. In fact, multiple analog RRAMs in parallel are required to obtain sufficient precision in weight storage to enable software equivalent learning [26][27]. As an alternative to RRAMs, Flash memory-based synapses have been proposed [28-34]. The channel hot electron injection (CHI) for programming during LTP [28-30] uses a large current to inject electrons in the floating gate, resulting in a high energy-loss [35]. NBTI based trap generation in high $\kappa$ dielectric MOSFETs is also shown to mimic synaptic activity, but it requires large electrical stressing to prepare the device [31-35]. Thus, a highly energy efficient, scalable synaptic device with gradual LTP and LTD is still challenging.

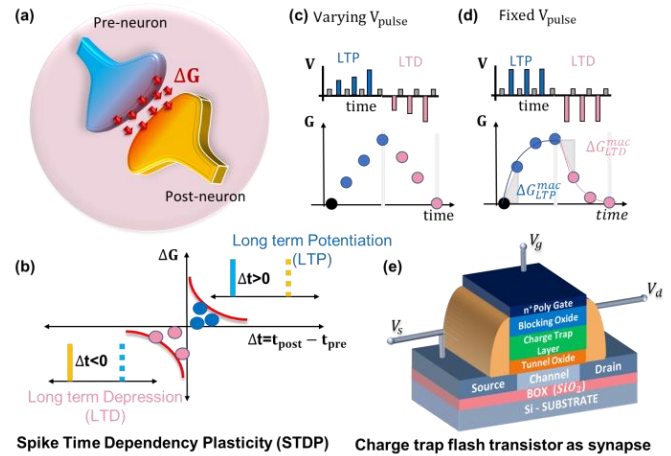

Fig. 1 **a)** A synapse connects a pre-neuron to a post-neuron. **b)** Synaptic conductance change ($\Delta G$) depends on the spike time difference ($\Delta t$) between post-neuron ($t_{post}$: yellow) and pre-neuron ($t_{pre}$: blue) known as Spike Time Dependent Plasticity (STDP). In memristors, write pulses (LTP: blue and LTD: pink) and read pulses (grey) are applied to change $G$. **c)** $G$ changes with the $V_{pulse}$ increase is typical memristors behavior. **d)** $G$ change and then saturates when fixed $V_{pulse}$ is applied repeatedly. The $\Delta G^{max}$ occurs initially. This is the critical requirement for SNNs **e)** An artificial synapse is implemented in a Charge Trap Flash (CTF) enables LTP & LTD by the threshold voltage ($V_T$) shift to replicate the learning rate behavior at fixed gate pulses.

In this paper, we propose a synapse based on highly manufacturable Charge Trap Flash (CTF) Memory (1F) device [36] on an SOI platform shown in Fig. 1e. The bulk Si technology based CTF capacitors exhibit gradual $V_T$ shift by Fowler Norheim (FN) tunneling to enable symmetric LTP & LTD with high energy efficiency and a designable learning rate. A careful design of operation and sub-circuit using SOI-platform and diodes (2D) in SPICE show the essential energy efficiency ($< 3\ fJ$/spike) is preserved by preventing undesirable (parasitic) leakage currents. A mathematical model of the experimental LTP / LTD is developed. A Spiking Neural Network (SNN) for Iris classification is implemented with CTF synpase to show excellent performance due to gradual LTP/LTD. Finally, a benchmarking of this work with the state-of-the-art is presented.

**Charge Trap Flash (CTF) device**

To demonstrate synapse by FN tunneling in CTF device, a $100\ \mu m \times 100\ \mu m$ CTF capacitor shown in Fig. 1e is fabricated as described in detail earlier [36]. In brief, the device is fabricated on an n-Si substrate with 4 nm thermal $SiO_2$ as the tunnel oxide, 6 nm LPCVD $Si_3N_4$ as the charge trap layer (CTL), and 12 nm MOCVD $Al_2O_3$ as the blocking oxide and n+ polysilicon on 300 mm Si substrate by Applied Materials cluster tool. Aluminum is used as the back contact. A self-aligned Boron implant provides a source for minority carriers for fast programming. In this CTF capacitor test vehicle, the bias is applied at the gate, and the source/drain and body are shorted to ground to measure total current. Though the demonstration is based on p-channel based CTF device, the conclusions are valid for n-channel CTF MOSFETs.

**Fowler Nordheim (FN) Tunneling based CTF Synapse**

The gradual program/erase operation to enable LTP & LTD is performed through FN tunneling, which is an electric field driven current transport to charge/discharge the CTL, which enables low current operation. The detailed physical mechanisms of CTF memory operation is presented elsewhere [37]. However, in brief, when a positive (negative) voltage is applied to the gate ($V_G$) with the body and source/drain grounded, electrons tunnel from (to) the channel through 4 nm tunnel oxide from (to) the CTL to enable threshold voltage shift $\Delta V_T > 0$ i.e., programming, $\Delta V_T < 0$ i.e. erase). The threshold voltage ($V_T$) translates to drain current ($I_D$), which represents synaptic conductance (G) as follows:

$$I_D = K(V_{GS} - V_T)V_{DS} \quad (1)$$
$$I_D = GV_{DS}$$
$$G = K(V_{GS} - V_T) \quad (2)$$

where $K$ is proportionality constant [38]. Erasing ($\Delta V_T < 0$) results in LTP ($\Delta G > 0$) and programming ($\Delta V_T > 0$) implies the LTD ($\Delta G < 0$). Thus, $V_T$ shift translates to conductance change ($\Delta G$) to enable LTP and LTD. Given that $V_T$ has a range from $V_{T-min}$ to $V_{T-max}$, and we applied $V_{GS} = V_{T-max}$; then,

$$G = K(V_{T-max} - V_T) \quad (3)$$

This implies a $G_{min} \approx 0$ and $G_{max} = K(V_{T-max} - V_{T-min})$ corresponding to $V_{T-max}$ and $V_{T-min}$.

Experimental LTD and LTP based on $V_T$ shift with repeated application of identical $V_G$ pulse is shown in Fig. 2a. The program pulse (12.5V for 1ms) and erase pulse (-14.5V for 20ms) are chosen such that symmetric operation LTD & LTP ($V_T$ shift from -1.3 V to -0.3 V, i.e., $Range(V_T) = 1\ V$) (same LTP and LTD window) occurs within 1000 pulses. As pulse voltage reduces for fixed pulse-width, the $V_T$ shift magnitude reduces. The $Range(V_T)$ after a fixed number of (say 1000) pulses for different pulse bias is plotted in Fig. 2b. The $Range(V_T)$ vs. $V_G$ curves are linearly extrapolated to $Range(V_T) = 0$, to estimate the threshold of $V_G$ for finite LTP and LTD. For 1000 pulses, the threshold was estimated as 9.8 V at 1 ms pulse width of LTD and -11.5 V at 20 ms pulse width of LTP. The $V_T$ change vs. programming time at fixed programming voltage ($V_G$) is well-documented [41].

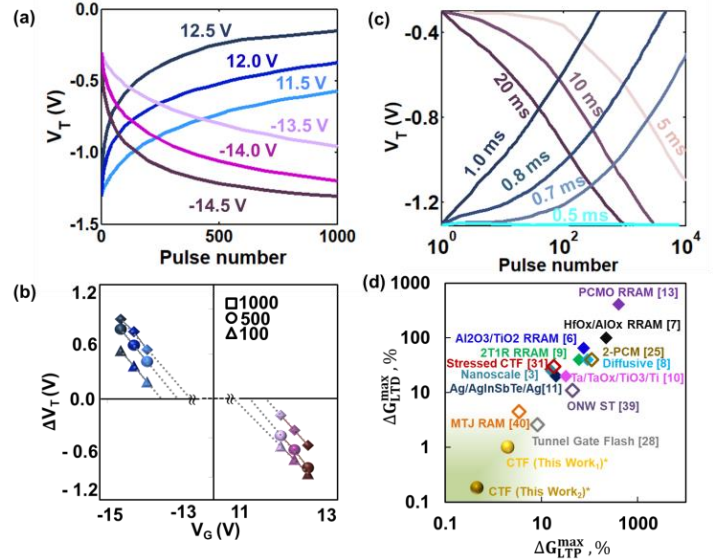

Fig. 2 **a)** Repeated application of same $V_{pulse}$ on gate ($V_G$) shows gradual $V_T$ shift for symmetric LTP and LTD with pulse-widths of 1 ms and 20 ms respectively. Higher $V_{pulse}$ increases $V_T$ range. **b)** Range of $V_T$ shift ($Range(V_T)$) extracted increases with the total number of pulses and $V_G$ respectively. The extrapolation to $Range(V_T) = 0$ estimates the threshold for write and erase for different pulse numbers. **c)** LTP and LTD behaviour can be tuned by the pulse-width arbitrarily, for fixed $V_G$ 12.5 V (LTD) and -14.5 V (LTP). **d)** A learning rate i.e. $\Delta G_{LTP}^{max}$ and $\Delta G_{LTD}^{max}$ normalized by range ($G_{max} - G_{min}$) is 10× lower for CTF compared to the state-of-the-art (filled squares are RRAM technology).

The gradual $V_T$ shift with pulse number can be designed alternatively by varying the pulse-width ($t_p$). Fig. 2c shows $V_T$ shift become more gradual with decreasing pulse-width for a fixed pulse amplitude (12.5 V for LTD & -14.5 V for LTP). Experimentally, $10^4$ states for LTP and LTD are demonstrated by pulse-width reduction, which is $10^2 \times$ improved than Flash [28-34] or memristor devices [4].

STDP has been demonstrated on several devices as synapse present in the literature [3][6-11][13][24-34]. Since the conductance change is non-uniform, the maximum conductance change ($|\Delta G_{LTP}^{max}| = |\Delta G_{LTD}^{max}|$) over the synaptic conductance-range ($G_{max} - G_{min}$) range should be a reasonable metric. The learning rate should be lower than 1-2% for supervised learning for analog-valued dataset [20-22][27]. As Fig. 2d shows, none of the other devices demonstrated as synapse in the literature have achieved this specification. Such a specification is satisfied for the first time. The gradual $V_T$ shift from Fig. 2a (for -13.5 V, 20 ms for LTP and 11.5 V, 1 ms for LTD) is made more gradual by pulse width based designability in Fig. 2c (-14.5 V, 5 ms for LTP and 12.5 V, 0.7 ms for LTD) as shown in Fig. 2d cited as This work$_2$.

To briefly understand the mechanism behind the observations in Fig. 2, the $V_T$ shift occurs due to charge storage in the charge trap layer of $Si_3N_4$ because of the imbalance between (i) the tunnel-in current ($J_{in}$) through tunnel oxide being higher than the (ii) tunnel-out current ($J_{out}$) through the blocking oxide i.e. $J_{in} \gg J_{out}$. As charge is stored, the electric field reduces in the tunnel oxide and increases in blocking oxide simultaneously to reduce the difference between $J_{in}$ and $J_{out}$. Eventually, a steady-state situation occurs when $J_{in} \sim J_{out}$ and no further charge storage occurs causing $V_T$ to saturate. Pulse-width reduction limits the time-duration of the net charging current i.e. $J_{in} - J_{out}$ (Fig. 2c). Thus, the rate of $V_T$ shift with pulses can be arbitrary reduced by reducing the pulse-width. Such a simple field dependent mechanism is not available in memristive devices, where usually complex interplay of self-heating based thermal runaway and ionic transport enables abrupt SET process [42-43], in contrast to the requirement of gradual conductance changes.

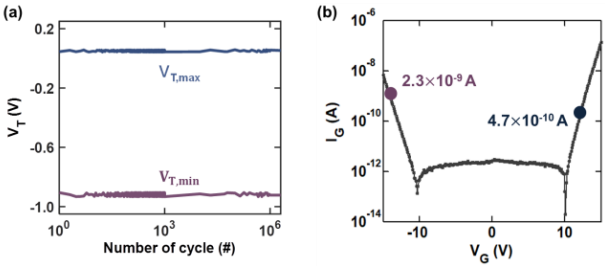

Fig. 3 **a)** Full $V_T$ swing for the window of Range($V_T$) ≈ 1V based endurance of more than $10^6$ cycle with low variability i.e., the coefficient of variation ($\sigma/\mu$) of 0.9% is demonstrated. It takes 1000 cycles for Range($V_T$) ≈ 1 of LTP and LTD during operation. **b)** Tunnelling current measured at the gate of the $100 \times 100$ $\mu m^2$ device shows very low current during programming/erasing.

The endurance of more than $10^6$ cycle for the window $\Delta V_T$ of 1V for the pulse of 12.0 V, 250 ms to -14.0 V, 10 s is shown in Fig. 3a. This endurance is 3 orders higher than high voltage (18 V) endurance presented earlier [36]. Each pulse during the endurance test is 1000× longer than the short single pulses for analog programming to enable large single step Range($V_T$) = $1V$ as an extreme case. Thus, the $10^6$ cycle with large Range($V_T$) ≈ $1V$ can be equivalent to $10^9$ gradual LTP/LTD cycles. Further, the coefficient of variation ($\sigma/\mu$) of the states compared to a memory window on 1V is 0.9%, which is extremely low compared to RRAM [12-13].

**Energy Cost due to Gate Leakage in Flash Synapse**
The *essential* energy loss during FN based programming for LTD and erasing for LTP is the tunneling current through the gate stack. At maximum voltage i.e. 12.5 V for LTP and -14.5 V for LTD, the current is 0.47 nA and 2.34 nA respectively for $100 \times 100$ $\mu m^2$ as shown in Fig. 3b. For scaled device at 180 nm node, the current is in 0.47- 2.31 fA range estimated by area scaling of current transport [44]. Based on the pulse-width, this results in the very low energy dissipation i.e., in 5.64 – 646.80 aJ range. This is extremely low, compared to other synapses presented in the literature [45], and comparable to biological synapses, which use 10 fJ per synaptic events [45]. Thus, LTP and LTD operations enabled by FN tunnelling in a scaled CTF device produce extreme energy efficiency. The number of levels in the proposed synaptic device will be limited by the cycle-to-cycle and device-to-device variability in $V_T$ updates. The experimentally measured cycle-to-cycle (C2C) variability (noise) and device-to-device (D2D) variability were measured to be σ/Range < 0.1%, much smaller than maximum $\Delta V_T$ per pulse ($max(\Delta V_T)/Range$ = 2%).

**CTF Scaling Feasibility:** The scalability of CTF to 180 nm node is key to estimate energy and yet maintain gradual LTP / LTD. First, CTF is highly manufacturable technology in production [44][46]. Variability in Flash has two sources (i) Cell-to-Cell Interference, which is less than 0.1% at the 90nm node and reduces even further with the inter-cell spacing increase [47]. (ii) Number of electrons per cell is ≈ $10^6$ at $Range(V_T)$ =1V at 100nm node, which will be sufficient to ensure low variability, as even a small $\Delta V_T = 1mV$ is represented by a 1000 electrons. Thus, it is immune to few electron problem of sub-30nm scaled Flash. Thus, relaxed scaling to 180nm node is promising. A total $V_T$ range of 1.0V requires a stored electron density $2 \times 10^{12}/cm^2$. A 2% maximum $\Delta V_T$ per spike is about 20mV. As charge density is constant, the area scaling reduces number of charges ($n$) – which leads to fluctuation as coefficient of variation ($\sigma/\mu = 1/\sqrt{n}$). At 200 nm, the 20±4 electrons are stored for the intended 20mV $V_T$ shift i.e. a 22% coefficient of variation ($\sigma/\mu$) – which is an estimation of scaling limit for devices. Alternatively, the variability based limit on minimum number of electrons per spike could also limit the minimum timescale for pulse-width.

**Flash Synapse Performance in the Array: Minimizing Parasitic Energy Loss**
To evaluate whether the extreme energy efficiency of the device translates to similar gains in a network, we present the implementation of the synapse in a crossbar array. This scheme for 2-terminal RRAMs explained in detail earlier [4] is adapted for 3-terminal CTF devices here. The FN tunneling based charge storage in CTF device is the primary process for LTP and LTD. To enable STDP, a pre-neuronal waveform (Fig. 4a (blue)) is applied to the gate, and a post neuronal waveform (Fig. 4a (brown)) is applied on the drain. The superposition of the gate voltage ($V_G$) and the drain voltage ($V_D$) (essentially channel voltage) offset by $\Delta t$ produces a peak voltage ($V_{peak}$) that depends upon $\Delta t$ in both polarity and magnitude i.e. $V_{peak}(\Delta t)$ as shown in Fig. 4b. The conductance change is related to $V_T$ shift as Equation (3). As $\Delta V_T$ depends upon $V_{peak}(\Delta t)$ as shown in Fig 4c, hence, the pre- and post-neuronal waveforms cause $\Delta G$ to change with $\Delta t$ i.e., $\Delta G(\Delta t)$ to produce STDP. The gate waveform has a positive spike of $V_{pos} = 9.8$ V and gentle negative voltage variation with peak negative voltage $V_{neg}= -3.0$ V. Similarly, the drain has a voltage spike $V_{pos} =11.5$ V with a gentle negative voltage variation with peak negative voltage $V_{neg}= -2.7$ V. These voltage magnitudes of the individual gate and drain waveforms are selected such that they produce no conductance change on their own i.e., they are below the write/erase threshold in Fig. 2b. However, the superposition of these pulses produces a $\Delta t$ dependent peak voltage in the range of 9.8 V to 12.5 V for $\Delta t < 0$ and -11.5 V to -14.5 V for $\Delta t > 0$ as shown in Fig. 4c. The superposition of the gate voltage ($V_G$) and the drain voltage ($V_D$) (essentially channel voltage) offset by $\Delta t$ produces a peak voltage ($V_{peak}$) that depends upon $\Delta t$ in both polarity and magnitude i.e. $V_{peak}(\Delta t)$ as shown in Fig. 4c. The Fig. 4d shows the actual dependence of peak voltage ($V_{peak}$) on $\Delta t$. When this peak voltage is applied to the CTF synapse, it results in a $\Delta t$ dependent $\Delta V_T$. Since $\Delta V_T$ implies $\Delta G$ we obtain a particular $\Delta G$ corresponding to a specific $\Delta t$, which is the STDP as shown in Fig. 4e. Here, $\Delta G$ is the difference between initial conductance ($G_i$) before the applied pulse final conductance ($G_f$) afterwards. We set $G_i = G_{min}$ for LTP, while $G_i = G_{max}$ for LTD to show the maximum $\Delta G$ corresponding to the $\Delta t$, which is equivalent to $V_T$ shift with $V_{peak}$ as shown in Fig. 2a. Also, $\Delta G$ is normalized with respect to range i.e. $G_{max} - G_{min}$ i.e. $\Delta G = \Delta G/(G_{max} - G_{min})$.

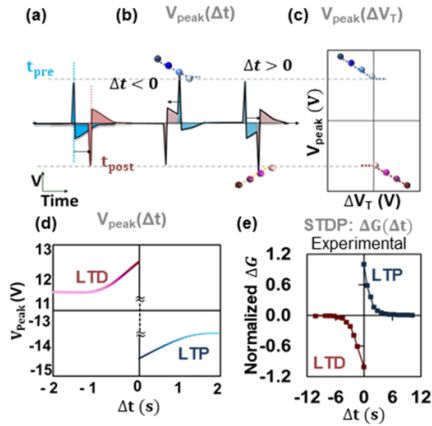

Fig. 4. **a)** Waveforms from pre-neuron at an instant $t_{pre}$ and post-neuron spike at instant $t_{post}$ staggered in time by $\Delta t = t_{post} - t_{pre}$. **b)** Superposition of pre- and post-neuron spike offset by time interval $\Delta t$ causes the resultant waveform to have a peak voltage ($V_{peak}$, brown and blue circles), whose sign and magnitude depends upon the sign and magnitude of $\Delta t$ i.e. $V_{peak}(\Delta t)$. **c)** The peak voltage causes $V_T$ shift i.e. $V_{peak}(\Delta V_T)$, which leads to conductance change $\Delta G(V_{peak})$. **d)** The actual dependence of peak voltage ($V_{peak}$) on $\Delta t$ is shown. **e)** Experimental STDP shows normalized $\Delta G$ changes with $\Delta t$ i.e. $\Delta G(\Delta t)$ achieved by applying pre- and post-neuronal waveforms on the CTF synapse.

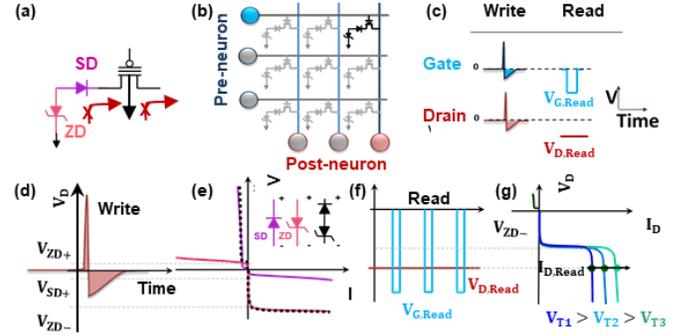

Fig. 5 **a)** Schematic of p-MOS based Flash synapse (1F) causes high $I_D$ from grounded source and body. Modified synapse where the source is connected to the ground via two diodes (2D) sub-circuit i.e. a standard $pn$ diode (SD) and Zener diode (ZD) in a back-to-back connection to produce the 1F2D configuration. **b)** The 1F2D synapses in a cross-bar array where the gate is connected to pre-neuron, the drain to post-neuron. **c)** Biasing scheme during (i) write operations and (ii) read operation. **d)** Post-neuron voltage waveform is compared to **e)** the IV characteristics of 2D sub-circuit (dashed) which blocks the current in the source-drain due to post-neuronal voltage waveform. Essentially, the IV characteristics of the SD (purple) and Zener diode (pink). **f)** Gate pulses and drain dc voltage during reading. **g)** $IV$ characteristics show that $I_D$ is limited by 2D sub-circuit which is off prior to Zener breakdown. However, during the read operation, the 2D-subcircuit turns on by Zener breakdown. Thus, the $I_D$ depends upon the $V_T$ of the Flash device.

The *essential* tunneling current through the gate (G) is extremely small when the body is grounded as shown in Fig. 3b. However, if the body (B) and the source (S) terminals are grounded as in a typical Flash device, then assuming that the channel is on (worst case), the post-neuronal waveform applied on the drain (D) terminal will drive a large, *non-essential* and undesirable drain current ($I_D$) as shown in Fig. 5a. $I_D$ has two components due to post-neuronal $V_D$ waveform - (i) **from grounded source** due to a $V_{DS}$ applied to a MOSFET in the on state in the worst-case situation. (ii) the forward bias current flowing **from the grounded body**. First, to cut off the drain to body junction forward bias leakage, CTF device is fabricated on an SOI substrate to produce a floating body *n*-channel MOSFET i.e., eliminate the body contact and any resultant leakage. Second, to eliminate the high S-D current due finite $V_{DS}$, a circuit element is used block current in the range of the drain bias due to the post-neuronal waveform. A simple and highly manufacturable solution of two diodes (2D) i.e., back-to back connected Zener diode with a standard $pn$ diode is added in series with SOI based CTF as shown in Fig. 5a. This 1F2D synapse is incorporated in a cross-bar network as shown in Fig. 5b. There are two modes of operation, write and read [48-49] as shown in Fig. 5c. In the write mode during learning, the write occurs by the superposition of gate and drain waveforms. In the read mode during inference, the read occurs when a dc drain bias is applied while the gate is pulsed with a read bias ($V_{G,read}$) when the pre-neuron spikes. When the post-neuronal bias (Fig. 5d) is applied for the **write operation**, the reverse bias $pn$ standard diode (SD) blocks the current for positive $V_D$, while the reverse biased Zener diode (ZD) blocks the current for negative $V_D$, as shown in Fig 5e. During the **read operation**, the 2D sub-circuit turns on as the read bias ($V_{D,read}$) exceeds $V_{ZD-}$ (Fig. 5f). Thus, the current is controlled by the Flash memory in series (Fig. 5g), which is dependent upon the $V_T$ of CTF. The 2D sub-circuit acts like a rectifier with a tunable threshold that depends on Zener breakdown voltage ($V_{ZD-}$) to block current and reduce parasitic energy loss during write but enable read when $V_D$ exceeds $V_{ZD-}$.

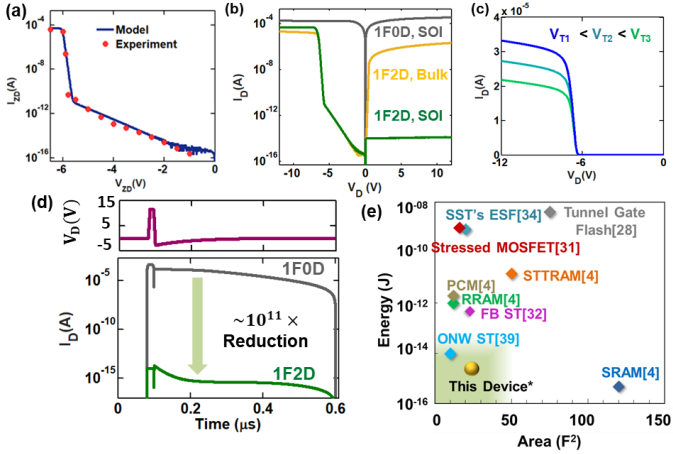

Fig. 6 **a)** Simulated IV characteristics of Zener diode (ZD) matched with Zener diode based on 180 nm GLOBALFOUNDRIES (GF) BCDLite CMOS technology. **b)** SPICE simulated $I_D - V_D$ characteristics for writing when CTF device is turned on (i.e. $V_G \gg V_T$) shows the worst case leakage estimate. Adding SOI (1F0D, SOI) reduces the body current. Adding the 2D- sub-circuit reduces (1F2D, SOI) reduces source current before Zener breakdown ($V_D < V_{ZD} \approx -6V$). **c)** $I_D - V_D$ characteristics shows that the $I_D$ depends on $V_T$ after Zener breakdown to enable read operation. **d)** During the write operation, the 1F2D, SOI synapse produces $\approx 10^{11} \times$ reduction in leakage compared to simply the CTF synapse (1F0D) assuming the transistor is on i.e. $V_G > V_T$ (worst case). This provides an ultra-low energy per write waveform of 2.5 fJ per spike, including the parasitic leakage. **e)** A comparison of this device (CTF) (scaled 180 nm) as synapse with state-of-the-art shows excellent footprint and energy performance. The energy performance competes with biology (10fJ)

To verify the idea, we simulate the 1F2D structure in NG-SPICE as shown in Fig. 6. While $pn$ diodes and Flash device IV characteristics (approximated by a TSMC 180 nm node technology transistor model) are well known, we chose a highly manufacturable Zener diode based on 180 nm GLOBALFOUNDRIES (GF) BCDLite CMOS technology with a Zener breakdown voltage, $V_{ZD-} = -6.0\,V$. The circuit model of a Zener diode was validated against the experimental IV as shown in fig. 6a. For the write operation, given that the Flash device is turned on ($V_G > V_T$) to represent the worst case, the SOI based CTF device only (1F0D-SOI) has high leakage due to the source to drain current for both positive and negative bias (Fig 6b). This is minimized by adding the 1F2D sub-circuit to enable the 1F2D (SOI) synapse. In case a bulk CTF device is used along with the 2 diodes, i.e. the 1F2D(bulk) synapse, then, in the positive bias, the drain-body MOSFET has a positive bias leakage if the body is grounded in bulk technology. This is mitigated by adding the Flash device (1F) on SOI technology as shown by the 1F2D (SOI) synapse. However, tunneling current flow primarily in the channel-gate direction is preserved, while other currents are blocked during the write operation. Thus, the 1F2D (SOI) synapse minimizes parasitic currents during the write operation as shown in Fig. 6b. For the read operation, simulated IV characteristics in Fig. 6c shows that 1F2D (SOI) synapse produces $V_T$ dependent $I_D$ for read bias when $V_D$ exceeds $V_{ZD-}$ when the gate voltage is above the threshold.

The energy estimation during the writing of a 1F2D (SOI) synapse by a post-neuronal waveform at a fixed gate bias is shown in Fig. 6d, where Flash device is turned on. The current in the 1F2D on SOI synapse is reduced by 11 orders of magnitude compared to the 1F0D synapse on bulk Si technology as shown in Fig. 6d. The energy including parasitic is 2.5 fJ, which is a slight enhancement over the 660 aJ of essential energy. Instead of the 2D sub-circuit, there are other possibilities like punch-through diodes [50] that can have designable asymmetry [51-52].

A CTF device has a layout area of $10F^2$, while the standard diode has $4F^2$ while zener diode has an area of $10F^2$ to produce a $< 30F^2$ synapse area. The general principles of peripheral circuit-design to supply high operational voltages akin to Flash memories is well-known [53]. Specific circuits needs to be developed to estimate systems level performance. The area vs. energy comparison of 1F2D (SOI) synapse with state-of-the-art shows excellent area and energy performance in Fig. 6e. In fact, it even competes with energy estimates of the biological synapse of 10fJ. Further, the input of synapse is high impedance. Given a spiking rate of 1kHz the impedance for 500nm CTF device is $10^{12}$ Ω which is significantly higher impedance compared to typical $10^6 Ω$ RRAMs. Hence it should support high fan-out as opposed to 2-terminal devices, which may load the source i.e., pre-neuron and also produces interconnect drop [54-55].

**Mathematical Model of STDP by Flash Synapse**

A modified spike time dependent plasticity rule, which incorporates weight dependent plasticity is used for learning [56]. The CTF flash is modeled for above mention SNN by mapping the time difference ($\Delta t$) between pre, and post synaptic spike ($t_{pre}$ and $t_{post}$) to peak voltage ($V_{peak}$) at the gate terminal by the equation given below:

$$V_{peak,Norm} = \frac{V_{peak} - \max(V_{peak})}{\max(V_{peak}) - \min(V_{peak})}$$

$\Delta t_{LTP} = -\tau_{LTP} \times V_{peak,Norm}$; $t_{LTD} = \tau_{LTD} \times V_{peak,Norm}$

Where, LTP, LTD is Long-Term Potentiation/Depression and $\tau_{LTP}, \tau_{LTD}$ : time constants for learning.

The conductance change across the flash device is modeled using two independent exponentially decaying functions ($f$ and $g$) with a maximum value of 1, accounting for spike time dependent plasticity and weight dependent plasticity respectively given by the equation below:

$$g_{LTP}(\Delta t_{LTP}) = e^{-a_1 \frac{G_i - G_{min}}{G_{max} - G_{min}}}$$

$$g_{LTD}(\Delta t_{LTD}) = e^{-a_2 \frac{G_{max} - G_i}{G_{max} - G_{min}}}$$

$$f_{LTP}(\Delta t_{LTP}) = e^{-\frac{\Delta t_{LTP}}{\tau_{LTP}}}$$

$$f_{LTD}(\Delta t_{LTD}) = e^{-\frac{\Delta t_{LTD}}{\tau_{LTD}}}$$

Where, $G_i$ is initial conductance. The device can undergo LTP or LTD, conductance change in both regimes is treated

separately using two different functions ($\Delta G_{LTP}$ and $\Delta G_{LTD}$) given below:

$$\Delta G_{LTP} = \Delta G_{LTP}^{max} \times f_{LTP}(\Delta t_{LTP}) \times g_{LTP}(G_i)$$
$$\Delta G_{LTD} = \Delta G_{LTD}^{max} \times f_{LTD}(\Delta t_{LTD}) \times g_{LTD}(G_i)$$

where $\Delta G_{LTD}^{max}$ and $\Delta G_{LTP}^{max}$ are the maximum conductance changes possible in one learning cycle.

**Flash synapse based SNN Performance**

We demonstrate supervised learning using a spiking neural network with mathematical modeling for a typical analog dataset – Fisher Iris flower classification [22] for CTF. The experimental data from the CTF based synapse is fit using the following equations:

$$\Delta G_{LTD} = \Delta G_{LTD}^{max} \times e^{-8.46\left(\frac{G_{max}-G_i}{G_{max}-G_{min}}\right)} \times e^{\frac{\Delta t_{LTD}}{\tau_{LTD}}}$$
$$\Delta G_{LTP} = \Delta G_{LTP}^{max} \times e^{-8\left(\frac{G_i-G_{min}}{G_{max}-G_{min}}\right)} \times e^{-\frac{\Delta t_{LTP}}{\tau_{LTP}}}$$

where $\Delta t_{LTD} = \Delta t < 0$ and $\Delta t_{LTP} = \Delta t > 0$. The fit parameters ($G_{min} = 0$, $G_{max} = 1$, $\Delta G_{LTD}^{max} = -0.14$, $\tau_{LTD} = 1.24\,s$ and, $\Delta G_{LTP}^{max} = 0.07$, $\tau_{LTP} = 1.05\,s$) produce an excellent match as shown in Fig. 7a-b. These synaptic models are incorporated into the network to be compared with ideal synapses described in the literature [44]. In Fig. 7c, learning produces an evolution of weights and consequent increase in classification for different initial weight randomizations (light blue) where the average behavior (black dashed) such that the performance converges with the "ideal" synapse to produce software-equivalent learning. To ensure this, at least 256 levels are needed [7], which is available for the CTF synapse. The noise in $V_T$ is $22 \times$ smaller than the resolution (i.e. 2% of maximum $V_T$ shift) and about $0.9 \times 10^{-3}$ times the range of $V_T$ and thus has no effect on the maximum learning accuracy achieved in the network.

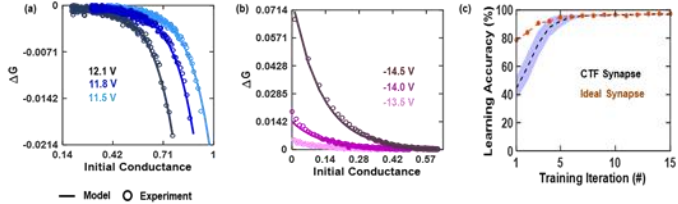

Fig. 7 Experimental **a)** LTD and **b)** LTP compared to the mathematical model (by curve fitting) for various write pulses. **c)** The experimentally validated synaptic model is implemented in SNN algorithm to compare with ideal synapse. Various random initialization of synaptic weights (light blue) with average (black dashed) behavior shows eventual convergence to software-equivalent learning performance (red).

**Benchmarking**

Finally, we benchmark our work against the state-of-the-art artificial synapses present in the literature in table 1. The CTF based synapse consumes very low energy. In fact, it is among the lowest energy possible after SRAM technology, which is volatile and expensive in terms of area (>120 $F^2$), binary and not amenable to cross-bar implementation. Further, low energy ONWST is based on an organic material flash not compatible with CMOS technology. Area of the CTF based synapse is highly competitive. The timescale is comparable to biology (~1 ms), which can be interesting for some real-time learning applications from natural data – as opposed to accelerated applications in software. This technology is highly manufacturable among CMOS silicon industry. The endurance is more than $10^9$ LTP/LTD cycle essential for multiple learning and reading cycle. This technology shows the gradually $10^4$ level of learning, which is a record improvement by 2 orders. Thus, the energy of write, CMOS compatibility, technological maturity, gradual and symmetric LTP and LTD, and LTP/LTD cycling reliability are the significant advantages.

"Table 1: Benchmarking with State-of-the-art"

| Synapse Technology | Energy (fJ) | Area ($F^2$) | Timing (ns) | CMOS Compatible | LTP/LTD cycles | Learning level | Gradual & Symmetric |
|---|---|---|---|---|---|---|---|
| SRAM[4] | 0.5 | 50 | 1 | High | ∞ | 2 | No |
| STTRAM[4,7] | >$10^4$ | 20 | >2 | Mid | $10^{12}$ | ~100 | No |
| PCM[4,24-25] | >$10^3$ | 4 | 50 | High | $10^{12}$ | >100 | No |
| RRAM[3-5] | >$10^4$ | 4 | >10 | High | $10^{11}$ | ~100 | No |
| ONW ST[39] | 1 | 8 | $10^4$ | Low | $10^4$ | 20 | **Yes** |
| Tun. gate Flash[28-30] | >$10^5$ | 75 | $10^4$ | High | $10^4$ | 1000 | **Yes** |
| SST's ESF[34] | >$10^3$ | 20 | >$10^4$ | High | NR | 100 | yes |
| Stressed MOSFET[31] | $10^5$ | 8 | >$10^4$ | High | NR | ~200 | No |
| FB ST[32-34] | >$10^3$ | 8 | $10^3$ | High | NR | 20 | Yes |
| **This work*** | **2.5** | **30** | **$10^5$** | **High** | **>$10^6$** | **~$10^4$** | **Yes** |

*energy estimated at a scaled 180 nm technology

**Conclusion**

In summary, we have proposed a synapse based on highly manufacturable Charge Trap Flash Memory on SOI technology. Gradual symmetric LTP and LTD is demonstrated where the number of states is arbitrarily tunable, e.g., shown as $10^1 - 10^4$ states by pulse characteristics. STDP is demonstrated by FN tunneling, which enables ultra-low energy dissipation (< 660 aJ) of essential energy. A 1F2D on SOI technology based synapse is presented to leverage the low energy operation in a synaptic array with a slight increase in total energy to 2.5 fJ energy per spike of LTP/LTD. The synaptic unit cell size of $30F^2$ is highly competitive. The synapse is highly reliable as it is capable of significant LTP/LTD cycles (> $10^6$ cycles of 1V) without any degradation. The experimentally measured noise in $V_T$ is $22 \times$ smaller than the resolution (i.e. 2% of maximum $V_T$ shift) and about $0.9 \times 10^{-3}$ times the range of $V_T$. These features compare very favorably against literature benchmarking. The record learning rate of 1-2% – a key specification to learn analog dataset– which is highly challenging for other synapse candidates in literature. An SNN using the CTF based synapse is used to solve a Fisher Iris classification problem with software equivalent performance. Such a synaptic array is a key step in enabling large scale neural network to mimic the human brain – including the timescale and energy.

**Acknowledgment**

The authors wish to acknowledge helpful discussions regarding flash memory with Mr. Sunny Sadana, the Research Associate at IIT Bombay. This work was supported in part by Nano Mission & Science and Engineering Research Board (SERB), the Department of Science and Technology (DST), Department of Electronics and IT, Govt. of India (DEITY).

**References**

1. P. Merolla, J. Arthur, F. Akopyan, N. Imam, R. Manohar, and D. S. Modha, "A digital neurosynaptic core using embedded crossbar memory with 45pJ per spike in 45nm," in *IEEE Custom Integrated Circuits Conference (CICC),* Sep 2011, pp. 1-4, doi: 10.1109/CICC.2011.6055294
2. T. Pfeil, T. C. Potjans, S. Schrader, W. Potjans, J. Schemmel, M. Diesmann, and K. Meier, "Is a 4-bit synaptic weight resolution enough?–constraints on enabling spike-timing dependent plasticity in neuromorphic hardware," in *Frontiers in neuroscience*, vol. 6, pp. 90, Jul 2017, doi: 10.3389/fnins.2012.00090
3. S. H. Jo, T. Chang, I. Ebong, B. B. Bhadviya, P. Mazumder, and W. Lu, "Nanoscale memristor device as synapse in neuromorphic systems," in *ACS Nano letters*, vol. 10, no. 4, pp. 1297-1301, Mar 2010, doi: 10.1021/nl904092h
4. J. J. Yang, D. B. Strukov, and D. R. Stewart, "Memristive devices for computing," in *Nature nanotechnology*, vol. 8, no. 1, pp. 13, Dec 2012, doi: 10.1038/nnano.2012.240
5. J. W. Jang, S. Park, Y. H. Jeong, and H. Hwang, "ReRAM-based synaptic device for neuromorphic computing," in *IEEE Circuits and Systems (ISCAS),* Jun 2014, pp. 1054-1057, doi: 10.1109/ISCAS.2014.6865320
6. M. Prezioso, F. M. Bayat, B. Hoskins, K. Likharev, and D. Strukov, "Self-adaptive spike-time-dependent plasticity of metal-oxide memristors," in *Nature Scientific Reports*, vol. *6*, pp. 21331, Feb 2016, doi: 10.1038/srep21331
7. S. Yu, D. Kuzum, and H. S. P. Wong, "Design considerations of synaptic device for neuromorphic computing," in *IEEE Circuits and Systems (ISCAS),* Jun 2014, pp. 1062-1065, doi: 10.1109/ISCAS.2014.6865322
8. Z. Wang, S. Joshi, S. E. Savel'ev, H. Jiang, R. Midya, P. Lin, M. Hu, N. Ge, J. P. Strachan, Z. Li, and Q. Wu, "Memristors with diffusive dynamics as synaptic emulators for neuromorphic computing," in *Nature materials*, vol. 16, no. 1, pp. 101, Sep 2016, doi: 10.1038/nmat4756
9. Z. Wang, S. Ambrogio, S. Balatti, and D. Ielmini, "A 2-transistor/1-resistor artificial synapse capable of communication and stochastic learning in neuromorphic systems," in *Frontiers in neuroscience*, vol. *8*, pp. 438, Jan 2015, doi: 10.3389/fnins.2014.00438
10. Y. F. Wang, Y. C. Lin, I. T. Wang, T. P. Lin, and T. H. Hou, "Characterization and modeling of nonfilamentary Ta/TaO x/TiO 2/Ti analog synaptic device," *Nature Scientific reports*, vol. 5, pp. 10150, May 2015, doi: 10.1038/srep10150
11. Y. Li, Y. Zhong, J. Zhang, L. Xu, Q. Wang, H. Sun, H. Tong, X. Cheng, and X. Miao, "Activity-dependent synaptic plasticity of a chalcogenide electronic synapse for neuromorphic systems," in *Nature Scientific reports*, vol. 4, pp. 4906, May 2014, doi: 10.1038/srep04906
12. S. Mandal, A. El-Amin, K. Alexander, B. Rajendran, and R. Jha, "Novel synaptic memory device for neuromorphic computing," in *Nature Scientific reports*, vol. 4, pp. 5333, Jun 2014, doi: 10.1038/srep05333
13. N. Panwar, D. Kumar, N. K. Upadhyay, P. Arya, U. Ganguly, and B. Rajendran, "Memristive synaptic plasticity in Pr 0.7 Ca 0.3 MnO 3 RRAM by bio-mimetic programming," in *IEEE Device Research Conference (DRC),* Jun 2014, pp. 135-136, doi: 10.1109/DRC.2014.6872334
14. S. Lashkare, N. Panwar, P. Kumbhare, B. Das, and U. Ganguly, "PCMO-Based RRAM and NPN Bipolar Selector as Synapse for Energy Efficient STDP," in *IEEE Electron Device Letters*, vol. 38*, no*. 9, pp. 1212-1215, Jul 2017, doi: 10.1109/LED.2017.2723503
15. S. Yu, Y. Wu, R. Jeyasingh, D. Kuzum, and H. S. P. Wong, "An electronic synapse device based on metal oxide resistive switching memory for neuromorphic computation," in *IEEE Transactions on Electron Devices*, vol. 58, no. 8, pp. 2729-2737, Jun 2011, doi: 10.1109/TED.2011.2147791
16. S. Mandal, B. Long, A. El-Amin, and R. Jha, "Doped HfO 2 based nanoelectronic memristive devices for self-learning neural circuits and architecture," in *IEEE Nanoscale Architectures (NANOARCH),* Jul 2013*,* pp. 13-18, doi: 10.1109/NanoArch.2013.6623030
17. B. Rajendran, Y. Liu, J. S. Seo, K. Gopalakrishnan, L. Chang, D. J. Friedman, and M. B. Ritter, "Specifications of nanoscale devices and circuits for neuromorphic computational systems," in *IEEE Transactions on Electron Devices*, vol. 60m no. 1, pp. 246-253, Jan 2013 , doi: 10.1109/TED.2012.2227969
18. C. Li, D. Belkin, Y. Li, P. Yan, M. Hu, N. Ge, H. Jiang, E. Montgomery, P. Lin, Z. Wang, and W. Song, "Efficient and self-adaptive in-situ learning in multilayer memristor neural networks," in *Nature Communications*, vol. 9, no. 1, pp. 2385, Jun 2018, doi: 10.1038/s41467-018-04484-2
19. S. Yu, "Neuro-inspired computing with emerging nonvolatile memorys," in *Proceedings of the IEEE*, vol. 106, no. 2, pp. 260-285, Feb 2018, 10.1109/JPROC.2018.2790840
20. G. Q. Bi, and M. M. Poo, "Synaptic modifications in cultured hippocampal neurons: dependence on spike timing, synaptic strength, and postsynaptic cell type," in *Journal of neuroscience*, vol. 18, no. 24, pp. 10464-10472, Dec 1998, doi: 10.1523/JNEUROSCI.18-24-10464.1998
21. L. F. Abbott, and S. B. Nelson, "Synaptic plasticity: taming the beast," in *Nature neuroscience*, vol. 3, no. 11s, pp. 1178, Nov 2000, doi: 10.1038/81453
22. A. Biswas, S. Prasad, S. Lashkare, and U. Ganguly, "A simple and efficient SNN and its performance & robustness evaluation method to enable hardware implementation," Dec 2016, *arXiv preprint arXiv:1612.02233*.
23. S. Yu, Y. Wu, R. Jeyasingh, D. Kuzum, and H. S. P. Wong, "An electronic synapse device based on metal oxide resistive switching memory for neuromorphic computation," in *IEEE Transactions on Electron Devices*, vol. 58, no. 8, pp. 2729-2737, Aug 2011, doi: 10.1109/TED.2011.2147791
24. M. Suri, O. Bichler, D. Querlioz, O. Cueto, L. Perniola, V. Sousa, D. Vuillaume, C. Gamrat, and B. DeSalvo, "Phase change memory as synapse for ultra-dense neuromorphic systems: Application to complex visual pattern extraction," in *IEEE Electron Devices Meeting (IEDM),* Dec 2011, pp. 4-4, doi: 10.1109/IEDM.2011.6131488
25. O. Bichler, M. Suri, D. Querlioz, D. Vuillaume, B. DeSalvo, and C. Gamrat, "Visual pattern extraction using energy-efficient "2-PCM synapse" neuromorphic architecture," in *IEEE Transactions on Electron Devices*, vol. 59, no. 8, pp. 2206-2214, Aug 2012, doi: 10.1109/TED.2012.2197951
26. I. Boybat, M. L. Gallo, S. R. Nandakumar, T. Moraitis, T. Parnell, T. Tuma, B. Rajendran, Y. Leblebici, A. Sebastian, and E. Eleftheriou, "Neuromorphic computing with multi-memristive synapses," in *Nature communications*, vol. 9, no. 1, pp. 2514, Jun 2018, doi: 10.1038/s41467-018-04933-y
27. A. Shukla, S. Prasad, S. Lashkare, and U. Ganguly, "A case for multiple and parallel RRAMs as synaptic model for training SNNs," Mar 2018 , *arXiv preprint arXiv:1803.04773*.
28. C. Diorio, P. Hasler, A. Minch, and C. A. Mead, "A single-transistor silicon synapse," in *IEEE transactions on Electron Devices*, col. 43, no. 11, pp. 1972-1980, Nov 1996, doi: 10.1109/16.543035
29. C. Diorio, P. Hasler, A. Minch, and C. A. Mead, "A floating-gate MOS learning array with locally computed weight updates," in *IEEE Transactions on Electron Devices*, vol. 44, no. 12, pp. 2281-2289, Dec 1997, doi: 10.1109/16.644652
30. H. S. Choi, D. H Wee, H. Kim, S. Kim, K. C. Ryoo, B. G. Park, and Y. Kim, "3-D Floating-Gate Synapse Array With Spike-Time-Dependent Plasticity," in *IEEE Transactions on Electron Devices*, vol. 65, no. 1, pp. 101-107, Jan 2018, doi: 10.1109/TED.2017.2775233
31. X. Gu, and S. S. Iyer, "Unsupervised Learning Using Charge-Trap Transistors," in *IEEE Electron Device Letters*, vol. 38, no. 9, pp. 1204-1207, Sep 2017, doi: 10.1109/LED.2017.2723319
32. H. Kim, J. Park, M. W. Kwon, J. H. Lee, and B. G. Park, "Silicon-based floating-body synaptic transistor with frequency-dependent short-and long-term memories," in *IEEE Electron Device Letters*, vol. 37, no. 3, pp. 249-252, Mar 2016, doi: 10.1109/LED.2016.2521863
33. H. Kim, S. Hwang, J. Park, and B. G. Park, "Silicon synaptic transistor for hardware-based spiking neural network and neuromorphic system," *Nanotechnology*, vol. 28, no. 40, pp. 405202, Oct 2017, doi: 10.1088/1361-6528/aa86f8
34. F. M. Bayat, X. Guo, H. A. Om'mani, N. Do, K. K. Likharev, and D. B. Strukov, "Redesigning commercial floating-gate memory for analog computing applications," in *IEEE Circuits and Systems (ISCAS)*, May 2015, pp. 1921-1924, doi: 10.1109/ISCAS.2015.7169048

35. Z. Chen, K. Hess, J. Lee, J. W. Lyding, E. Rosenbaum, I. Kizilyalli, S. Chetlur, and R. Huang, "On the mechanism for interface trap generation in MOS transistors due to channel hot carrier stressing," in *IEEE Electron Device Letters*, vol. 21, no. 1, pp. 24-26, Jan 2000, doi: 10.1109/55.817441
36. C. Sandhya, U. Ganguly, N. Chattar, C. Olsen, S. M. Seutter, L. Date, R. Hung, J. M. Vasi, and S. Mahapatra, "Effect of SiN on performance and reliability of charge trap flash (CTF) under Fowler–Nordheim tunneling program/erase operation," in *IEEE Electron Device Letters*, vol. 30, no. 2, pp. 171-173, Feb 2009, doi: 10.1109/LED.2008.2009552
37. H. Bachhofer, H. Reisinger, E. Bertagnolli, and H. Von Philipsborn, "Transient conduction in multidielectric silicon–oxide–nitride–oxide semiconductor structures," in *Journal of Applied Physics*, 89(5), pp. 2791-2800, Feb 2001, doi: 10.1063/1.1343892
38. Taur, Yuan, and Tak H. Ning. *Fundamentals of modern VLSI devices*. Cambridge university press, 2013, doi: 10.1017/CBO9781139195065
39. W. Xu, S. Y. Min, H. Hwang, and T. W. Lee, "Organic core-sheath nanowire artificial synapses with femtojoule energy consumption," in *Science advances*, vol. 2, no. 6, pp. 1501326, Jun 2016, doi: 10.1126/sciadv.1501326
40. P. Krzysteczko, J. Münchenberger, M. Schäfers, G. Reiss, and A. Thomas, "The Memristive Magnetic Tunnel Junction as a Nanoscopic Synapse-Neuron System," in *Advanced Materials*, vol. 24, no. 6, pp. 762-766, Jan 2012, doi: 10.1002/adma.201103723
41. Bu, J., & White, M. H. (2001). Design considerations in scaled SONOS nonvolatile memory devices. *Solid-State Electronics*, 45(1), 113-120. https://doi.org/10.1016/S0038-1101(00)00232-X
42. N. Panwar, A. Khanna, P. Kumbhare, I. Chakraborty, and U. Ganguly, "Self-heating during submicrosecond current transients in Pr 0.7 Ca 0.3 MnO 3-based RRAM," in *IEEE Transactions on Electron Devices*, vol. 64, no. 1, pp.137-144, Jan 2017, doi: 10.1109/TED.2016.2632712
43. I. Chakraborty, N. Panwar, A. Khanna, and U. Ganguly, "Space Charge Limited Current with Self-heating in $Pr_{0.7}Ca_{0.3}MnO_3$ based RRAM," May 2016, *arXiv preprint arXiv:1605.08755*.
44. C. Y. Lu, K. Y. Hsieh, and R. Liu, "Future challenges of flash memory technologies," in *Microelectronic engineering*, vol. 86, no. 3, pp. 283-286, Mar 2009, doi: 10.1016/j.mee.2008.08.007
45. D. Kuzum, S. Yu, and H. S. P. Wong, "Synaptic electronics: materials, devices and applications," *Nanotechnology*, vol. 24, no. 38, pp. 382001, Sep 2013, doi: 10.1088/0957-4484/24/38/382001
46. D. Kang, W. Jeong, C. Kim, D. H. Kim, Y. S. Cho, K. T. Kang, and H. Lee, "256 Gb 3 b/cell V-NAND flash memory with 48 stacked WL layers," in *IEEE Journal of Solid-State Circuits*, vol. 52, no. 1, pp. 210-217, Feb 2016, doi: 10.1109/ISSCC.2016.7417941
47. K. Prall, "Scaling non-volatile memory below 30nm," in *IEEE Non-Volatile Semiconductor Memory Workshop,* Aug 2007, pp. 5-10, doi: 10.1109/NVSMW.2007.4290561
48. A. Shukla, V. Kumar, and U. Ganguly, "A software-equivalent SNN hardware using RRAM-array for asynchronous real-time learning," in *IEEE Neural Networks (IJCNN),* May 2017, pp. 4657-4664, doi: 10.1109/IJCNN.2017.7966447
49. A. Shukla, and U. Ganguly, "An On-Chip Trainable and the Clock-Less Spiking Neural Network With 1R Memristive Synapses," *IEEE Transactions on Biomedical Circuits and Systems*, vol. 12, no. 4, August 2018, doi: 10.1109/TBCAS.2018.2831618
50. V. S. S. Srinivasan, S. Chopra, P. Karkare, P. Bafna, S. Lashkare, P. Kumbhare, Y. Kim, S. Srinivasan, S. Kuppurao, S. Lodha, and U. Ganguly, "Punchthrough-diode-based bipolar RRAM selector by Si epitaxy," in *IEEE Electron Device Letters*, vol. 33, no. 10, pp. 1396-1398, Oct 2012, doi: 10.1109/LED.2012.2209394
51. R. Mandapati, A. S. Borkar, V. S. Srinivasan, P. Bafna, P. Karkare, S. Lodha, and U. Ganguly, "The impact of npn selector-based bipolar RRAM cross-point on array performance," in *IEEE Transactions on Electron Devices*, vol. 60, no. 10, pp. 3385-3392, Oct 2013, doi: 10.1109/TED.2013.2279553
52. S. Lashkare, P. Karkare, P. Bafna, M. V. S. Raju, V. S. S. Srinivasan, S. Lodha, U. Ganguly, J. Schulze, and S. Chopra, "A bipolar RRAM selector with designable polarity dependent on-voltage asymmetry," in *IEEE Memory Workshop (IMW),* May 2013, pp. 178-181, doi: 10.1109/IMW.2013.6582128
53. Micheloni, Rino, and Luca Crippa. "Charge pumps, voltage regulators and HV switches." *Inside NAND Flash Memories*. Springer, Dordrecht, 2010. 299-327, https://doi.org/10.1007/978-90-481-9431-5_11
54. S. B, Eryilmaz, D. Kuzum, S. Yu, and H. S. P. Wong, "Device and system level design considerations for analog-non-volatile-memory based neuromorphic architectures," in *IEEE Electron Devices Meeting (IEDM),* Dec 2015, pp. 4-1, doi: 10.1109/IEDM.2015.7409622
55. R. Mandapati, A. Borkar, V. S. S. Srinivasan, P. Bafna, P. Karkare, S. Lodha, B. Rajendran, and U. Ganguly, "On pairing of bipolar RRAM memory with NPN selector based on set/reset array power considerations," in *IEEE Transactions on Nanotechnology*, vol. 12, no. 6, pp. 1178-11, Nov 2013, doi: 10.1109/TNANO.2013.2284508
56. S. Prasad, A. Biswas, A. Shukla, U Ganguly, "A Highly Efficient Performance and Robustness Evaluation Method for a SNN based Recognition Algorithm," International Conference on Artificial Neural Networks (ICANN) 2017 (accepted), doi: arXiv:1612.02233v1